# Developing an NLP-based Recommender System for the Ethical, Legal, and Social Implications of Synthetic Biology[1]

Damien Dablain[*], Lilian Huang[**] and Brandon Sepulvado[***]

[*]*ddablain@nd.edu*
Department of Computer Science and Engineering, 384 Fitzpatrick Hall Notre Dame, IN 46556, USA

[**] *huang-lilian@norc.org*
Department of Statistics and Data Science, NORC at the University of Chicago, 55 East Monroe Street, 30$^{th}$ Floor, Chicago, IL 60603, USA

[***] *sepulvado-brandon@norc.org*
Department of Methodology and Quantitative Social Sciences, NORC at the University of Chicago, 4350 East-West Hwy 8th Floor, Bethesda, MD 20814, USA

**Introduction**
Synthetic biology is an emerging field that involves the engineering and re-design of organisms for purposes such as food security, health, and environmental protection. As such, it poses numerous ethical, legal, and social implications (ELSI) for researchers and policy makers. Various efforts to ensure socially responsible synthetic biology are underway. Policy making is one regulatory avenue, and other initiatives have sought to embed social scientists and ethicists on synthetic biology projects (Balmer et al., 2015). However, given the nascency of synthetic biology, the number of heterogeneous domains it spans (Raimbault et al., 2016), and the open nature of many ethical questions, it has proven challenging to establish widespread concrete policies, and including social scientists and ethicists on synthetic biology teams has met with mixed success.

This text proposes a different approach, asking instead is it possible to develop a well-performing recommender model based upon natural language processing (NLP) to connect synthetic biologists with information on the ELSI of their specific research? This recommender was developed as part of a larger project building a Synthetic Biology Knowledge System (SBKS) to accelerate discovery and exploration of the synthetic biology design space (Mante et al., 2021). Our approach aims to distill for synthetic biologists relevant ethical and social scientific information and embed it into synthetic biology research workflows.

**Data**
To develop such a recommender, we generated a corpus of 9,684 synthetic biology and 180 ELSI articles from PubMed Central. The corpus was created using the keyword-based query developed by Shapira et al. (2017). This query is for synthetic biology, so we identified ELSI texts within the resulting synthetic biology articles by further adding the following terms: ethical, ethics, bioethics*, policy, governance, biosafety, social issues, social impact, environmental impact, and environmental issues.

**Methods**
Step 1: Representing Synthetic Biology and ELSI Knowledge: We rely upon text data to capture synthetic biology and ELSI knowledge. Word embeddings are one of the most popular ways to represent language in a corpus. Although training custom embeddings for a given task requires large amounts of data, which would be prohibitive for a relatively new field such as synthetic biology, pre-trained embeddings exist. Word embeddings trained on one corpus can be used to represent knowledge in another corpus if the latter is sufficiently similar; this process is known as transfer learning. However, the quality of pre-trained embeddings for a given task or representing a specific corpus can be improved with fine-tuning, which entails updating the pretrained model for a prediction task using the corpus of interest.

---

[1] This material is based upon work supported by the National Science Foundation under Grant No. 1939887.

We investigate the performance of four pre-trained language models in representing synthetic biology knowledge. Bidirectional Encoder Representations from Transformers (BERT) (Devlin et al., 2019), a general corpus to use as our baseline; SciBERT, which is trained on 1.14M papers from Semantic Scholar (Beltagy et al., 2019), BioBERT, which is trained on PubMed abstracts, PubMed Central full-text articles, and English Wikipedia and BooksCorpus words, (Lee et al., 2020), and PubMedBERT, which is trained on 21 GB of PubMed abstracts (Gu et al., 2022). A key limitation of these language models is that none is focused on synthetic biology; even the scientific and biomedical models are still more general.

As such, we fine tune these models by predicting the topic of each synthetic biology paper. Doing so allows us to implicitly train embeddings that position documents about the same topic as closer together in embedding space than documents about different topics (Sun et al., 2020). However, these topics are not known beforehand. To identify topics discussed by each paper, we used Latent Dirichlet Allocation (LDA) topic models on the abstracts of synthetic biology papers. LDA identifies topics present in a corpus and the extent to which texts engage each topic. We identified five topics with LDA and confirmed the appropriateness of these topics by consulting multiple synthetic biologists on our team. LDA output scores each paper for each of the five topics, indicating the extent to which each abstract discusses the topics. For the purposes of fine-tuning, we labeled each abstract with a single topic by assigning the topic on which the abstract scored the highest. We then fine tune each model by predicting synthetic biology abstracts' topics. We train for 10 epochs, using the Adam optimizer and a .00002 learning rate.

The final part of Step 1 involves generating document representations for ELSI articles. To do so, we use the language models to encode each ELSI abstract. The four models output an embedding of a paper abstract in the form of word tokens. For example, given a batch of 10 abstracts, the model outputs a [10 X 512 X 768] matrix, where 10 is the batch size, 512 corresponds to the number of word tokens in an individual abstract[i] and 768 is the vector representation of a single word token. The models also output a single embedding each abstract in the batch (pooled output). We applied the Tanh activation function to the pooled output. A matrix of ELSI abstract embeddings organized by predicted topic is output from this stage.

Step 2: Generating ELSI Article Recommendations: Once the language models are fine-tuned and we have matrix of representations for ELSI abstracts, the necessary components for our recommender are in place. Figure 1 illustrates the process. A synthetic biology abstract is fed to the model, which generates an embedding and a topic prediction. This embedding is compared to the ELSI embeddings (matrix) for that same predicted synthetic biology topic article, and a recommended ELSI article is output. The recommended ELSI article is the paper that is closest in distance to the synthetic biology document. We use the L1 distance metric as our similarity measure.

Figure 1. Recommender Workflow.

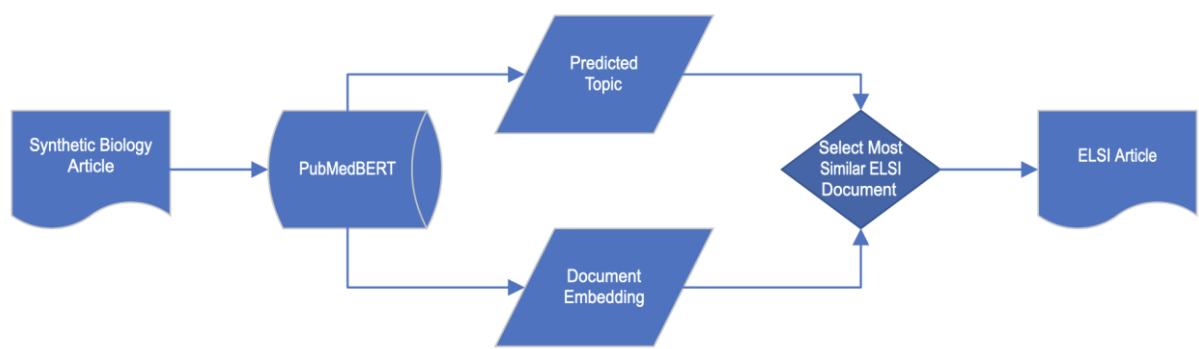

**Results and Ongoing Research**
Table 1 presents accuracy and F1 scores for each of the pre-trained models on the task of classifying the topics of synthetic biology texts. PubMedBERT was the best performing model. Since it performed best, we used PubMedBERT to generate ELSI recommendations. Because BERT is not domain specific (e.g., to STEM or biomedical fields), it is not surprising that it exhibited the worst performance. However, it is potentially surprising that PubMedBERT outperforms BioBERT because the corpus on which BioBERT was trained includes texts from PubMed Central, whereas PubMedBERT includes only PubMed abstracts.

Table 1. Performance of language models.

| Model | Accuracy | Macro F1 |
|---|---|---|
| BERT | 79.39 | 78.80 |
| SciBERT | 87.50 | 87.37 |
| BioBERT | 88.04 | 87.97 |
| PubMedBERT | 89.12 | 89.25 |

The second way we evaluated our model is with respect to the quality of recommendations. Our team manually reviewed recommendations for this component. Overall, the recommendations appear relevant. However, we noticed that several recommended articles include an extensive mention of the concerns that the research might raise without explicitly discussing ethical and/or policy resolution(s). Ongoing research will investigate two potential resolutions to this issue. First, the query to obtain relevant articles might be refined, and second, we will test adding an additional step in the workflow to predict ELSI mention versus engagement. With these additional steps, we aim to produce an even more useful tool that helps connect synthetic biologists with knowledge of the ELSI of their research.

---

[i] We restricted—for methodological purposes—the number of words in an abstract to 512.